\documentclass[12pt]{article}

\begin{document}
\vspace*{-.6in} \thispagestyle{empty}
\begin{flushright}
CALT-68-2287\\
CITUSC/00-041\\
hep-th/0007130
\end{flushright}
\baselineskip = 20pt

\vspace{.5in} {\Large
\begin{center}
Recent Progress in Superstring Theory\end{center}}

\begin{center}
John H. Schwarz\footnote{Work supported in
part by the U.S. Dept. of Energy under Grant No.
DE-FG03-92-ER40701.}
\\ \emph{California Institute of Technology\\
Pasadena, CA  91125 USA}
\end{center}
\vspace{1in}

\begin{center}
\textbf{Abstract}
\end{center}
\begin{quotation}
\noindent Superstring theory has continued to develop at a rapid
clip in the past few years. Following a quick review of some of
the major discoveries prior to 1998, this talk focuses on a few of
the more recent developments. The topics I have chosen to present
are 1) the use of K-theory to classify conserved charges carried
by D-branes; 2) tachyon condensation on unstable D-brane systems;
and 3) an introduction to noncommutative field theories and their
solitons.
\end{quotation}
\vspace{1in}

\centerline{Talk presented at the Conference {\it Quantization, Gauge Theory, and Strings}}
\centerline{dedicated to the memory of Professor Efim Fradkin}
\centerline{and at the {\it Ninth Marcel Grossmann Meeting} (MG9)} \vfil

\newpage

\pagenumbering{arabic} 

\section{Introduction}

The purpose of this talk is to survey some of the recent progress
in superstring theory that has occurred in the past few years in a
way that is accessible to nonexperts. A great deal has happened,
much more than can be covered in a 40-minute talk, so difficult
choices needed to be made.  Even for the topics that are included
it is necessary to sacrifice depth for breadth. All I hope to
achieve is to convey a general sense of the directions in which
the subject is developing and the impressive achievements that
have taken place.

The plan of this talk is as follows.  First I will give a quick
review of developments that took place prior to mid-1998.  Then I
will focus on three areas in which there has been important
progress since then.  The first of these concerns the
classification of conserved charges for D-brane systems.  As I
will explain, the appropriate mathematical category for this
purpose turns out to be K-theory.  Next I will discuss unstable
D-brane systems, which have one or more tachyon fields in the
world-volume theory.  We will describe the physics associated with
``tachyon condensation'' when the tachyon fields roll to a minimum
of the appropriate potential energy function. Finally, I will
introduce non-commutative geometry.  This turns out to be relevant
for D-brane systems that have a constant magnetic field in their
world volume.  Certain simplifying features associated with the
strong noncommutativity limit and their implications for string
theory will be discussed.

\section{Survey of Prior Results}

As of 1985, following some 15 years of development, five
consistent superstring theories had been identified \cite{GSW}.  Each of
these requires ten dimensions (nine spatial and one time
dimension), and each is supersymmetric.  The five theories are
shown in the table. The two type II theories have 32 conserved
supersymmetry charges, whereas each of the other three theories
have 16 conserved supercharges (corresponding to a Majorana--Weyl
spinor).

\begin{center}
\vspace{.2in}

\begin{tabular}{l|c|c}
&\textbf{Gauge Group} & \textbf{Chiral?}\\ \hline \textbf{Type I}
& $SO(32)$ & yes\\ \hline \textbf{Type IIA} & $U(1)$ & no\\ \hline
\textbf{Type IIB} & -- & yes\\ \hline \textbf{Heterotic} & $E_8
\times E_8$ & yes\\ \hline \textbf{Heterotic} & $SO(32)$ & yes\\
\hline
\end{tabular}
\end{center}
\begin{center}
{\bf Table 1.}  The Five Superstring Theories
\end{center}

Following another 10 years of development, and the ``second
superstring revolution,'' our understanding was significantly
enhanced \cite{JP}.  An important ingredient was the discovery of various
dualities that imply deep and subtle relationships among the five
theories.  In fact it became clear that they should best be viewed
as five distinct quantum vacua of a single underlying theory.

Even though the underlying theory is still not fully formulated,
it is clear that it is completely unique, with no adjustable
dimensionless parameters. Any dimensionless numbers that
characterize a particular quantum vacuum are determined by
expectation values of scalar fields. In particular, the string
coupling constant, $g_s$, is given by the expectation value of the
dilaton field (which belongs to the supergravity multiplet).  Each
of the five theories has an interesting limit as $g_s \rightarrow
\infty$.  The two SO(32) theories are S-dual, which means that
$g_s$ for one of them corresponds to $1/g_s$ for the other.  Thus,
when one theory is strongly coupled, it corresponds to the other
one at weak coupling. The IIB theory is self-dual in this sense.
Most remarkably, the remaining two theories --- type IIA and $E_8
\times E_8$ heterotic
--- turn out to have an 11th dimension whose size is proportional
to $g_s$, and thus is invisible in perturbation theory.  In the
limit $g_s \rightarrow \infty$ each of them gives an
11-dimensional vacuum, which is called ``M-theory.''  Defined in
this way, M-theory is neither more fundamental nor less
fundamental than the five superstring theories --- it is just one
more special limit in the moduli space of consistent quantum
vacua.  There is some confusion on this point, because the term
M-theory has also been used to refer to the underlying theory,
which is more fundamental.

Another important lesson of the second superstring revolution was
that in addition to the fundamental strings, the five superstring
theories (and M-theory) also have various nonperturbative objects
called $p$-branes.  These are dynamical structures which have $p$
extended spatial dimensions and can be idealized in many
situations as being infinitely thin in the $9 - p$ (or $10-p$)
normal directions. The basic examples of $p$-branes are ones that
carry a conserved $p$-brane charge that couples to a $(p +
1)$-form gauge field. Such a field can be regarded as a
generalized Maxwell field with $p+1$ antisymmetrized indices and
represented as a differential form
\begin{equation}
A^{(p+1)} = A_{\mu_{1}\mu_{2}\ldots \mu_{p+1}} dx^{\mu_{1}} \wedge dx^{\mu_{2}}
\wedge\ldots \wedge dx^{\mu_{p+1}}.
\end{equation}
The fact that the $p$-brane acts as a source for such a gauge field, with
charge $q$, is represented by the interaction term
\begin{equation}
S_{int} = q \int_Y A^{(p + 1)}.
\end{equation}
Here $A^{(p+1)}$ represents the pullback to the $p$-brane
world-volume $Y$. Such a $p$-brane is supersymmetric, preserving
half of the supersymmetry of the vacuum in which it is embedded,
provided that the BPS condition,
\begin{equation}
q = T_p \, ,
\end{equation}
is satisfied.  Here $T_p$ denotes the tension of the $p$-brane, which is its
energy density.

Let me list the BPS $p$-branes of the type II theories.  First of
all, these theories contain a two-form gauge field $B^{(2)}$ (in
the NS-NS sector).  The fundamental string (F1) is an electrical
source of this field.  Its magnetic dual, which is a source for
the dual six-form potential, is called an NS5-brane. In addition,
the type II theories have a collection of gauge fields $C^{(p+1)}$
(in the $R-R$ sector), for which a class of $p$-branes, called
Dirichlet-branes (or D$p$-branes) are sources. The $p$ values that
occur are even for the IIA theory and odd for the IIB theory.

The tensions of the various $p$-branes mentioned above are exactly determined
by the BPS condition.  Aside from a power of $2\pi$, the results (in string
metric) are as follows:
\begin{eqnarray}
T_{F1} &=& m_s^2\nonumber \\
T_{Dp} &=& m_s^{p+1}/g_s\nonumber \\
T_{NS5} &=& m_s^6/g_s^2.
\end{eqnarray}
Here, $m_s$ denotes the fundamental string mass scale.  It is
related to the fundamental string length scale $\ell_s$ and the
universal Regge slope $\alpha'$ by $m_s = \ell_s^{-1}$ and
$\alpha' = \ell_s^2$.  Note that the D-branes and the NS5-branes
are both non-perturbative, but that the D$p$-branes are much
lighter at weak coupling, so that their non-perturbative effects
dominate in this limit.  The D-branes have played a central role
in superstring theory research ever since the appearance of the
breakthrough paper by Polchinski \cite{JPD}.

The Dirichlet branes are called such because they are
hypersurfaces $Y\subset X$ on which the fundamental string can
end. ($X$ denotes the 10-dimensional spacetime.)  Thus, at weak
string coupling, their dynamics is determined by open-string field
theory specialized to open strings that end on $Y$.  In
particular, let us consider the case of $N$ coincident D-branes
and focus on the massless states of the open strings, which are
the relevant ones at low energies. In this limit one finds $N^2$
massless vector bosons belonging to the adjoint representation of
a $U(N)$ gauge group. Altogether, including their superpartners,
one has a supersymmetric $U(N)$ gauge theory on $Y$.

The special case of coincident D$3$-branes in the type IIB theory
is especially interesting, because the associated ${\cal N} = 4$,
$U(N)$ gauge theory that gives the dynamics is superconformal.
This means, in particular, that the quantum theory is finite at
each order of perturbation theory, with vanishing $\beta$
function. When $N$ is large the back reaction of the branes on the
spacetime geometry becomes important, and one should consider the
gravitational field of the D$3$-branes. This gives a
higher-dimensional analog of a black hole with a horizon. In
particular, one finds that the near-horizon geometry is that of
$AdS_5 \times S^5$, which is part of an exact solution of type IIB
string theory.

These observations led Juan Maldacena in November 1997 to put
forward his famous conjecture \cite{JM}, which asserts that ${\cal N} = 4$,
$d = 4$ $U(N)$ gauge theory is dual to type IIB superstring theory
on $AdS_5 \times S^5$ with the following identifications: $$ g_s
\leftrightarrow g^2_{YM}$$ $$ R(AdS_5)/\ell_s = R(S^5)/\ell_s
\leftrightarrow (g_{YM}^2 N)^{1/4}$$
\begin{equation}
\int_{S^5} F^{(5)} \leftrightarrow N.
\end{equation}
Here I have omitted numerical factors.  $R(AdS_5)$ and $R(S^5)$
denote the radii of the respective spaces, and $F^{(5)}$ is the
self-dual RR 5-form field strength of the IIB theory. Altogether,
one has an equivalence of a conformally invariant field theory
(CFT) and string theory in an anti de Sitter geometry, so one
speaks of AdS/CFT duality.

In his paper, Maldacena also proposed several other equivalences
of the same general type: namely, conjectural equivalences between
conformal field theories in $p+1$ dimensions, associated to
coincident $p$-branes, and string theory or M-theory in a geometry
that contains an $AdS_{p+2}$ factor. The way the duality relates
CFT correlation functions to AdS amplitudes was explained in
papers by Gubser, Klebanov, and Polyakov \cite{GKP} and by Witten
\cite{W1} a few months later. This has become an enormous, and
very fruitful, subject. By now, the Maldacena paper has well over
1000 citations. There is little doubt that the conjectures are
correct. I will not say more about them here other than to point
you to an excellent review paper \cite{MAGOO}, in case you wish to
learn more.

\section{Classification of D-brane Charges}

Consider, as before, a D-brane system with world volume $Y$
embedded in a spacetime manifold $X$ of dimension $d = 10$
(superstrings) or $d = 26$ (bosonic strings).  The surface $Y$
will shrink to a minimal surface due to the brane tension
(radiating away any excess energy).  Even then, the D-brane may be
unstable unless it is protected by a conserved charge.  A signal
of instability is when the open string spectrum on $Y$ contains
one or more tachyonic modes. Such tachyonic modes occur for the
following cases, in particular:
\begin{enumerate}
\item All D-branes in the $d=26$ bosonic string theory
\item Type II D$p$-branes with a ``wrong'' $p$ value (i.e., $p$ odd in IIA or $p$
even in IIB)
\item Type II D-brane $+$ anti-D-brane systems.
\end{enumerate}
Lest the terminology cause confusion, let me emphasize that there
never are physical tachyons. Their apparent presence simply means
that the vacuum under consideration is unstable. As for the Higgs
fields in the standard model, one should like for a stable minimum
of the appropriate potential. In the remainder of this section I
will discuss the classification of conserved D-brane charges, and
in the next section I will discuss the fate of unstable D-brane
systems when the tachyon rolls to the minimum of its effective
potential.

Since, as we have already indicated, the RR gauge fields that
couple to D-branes can be represented as differential forms, it
would be natural to guess that D-brane charges are given by
cohomology classes.  In simple cases this is good enough, but for
some topologies it can give wrong answers.  The right answer, as I
will explain, is given by K-theory classes \cite{MM, W2}. This is important,
because K-theory and cohomology can have different torsion groups.

To understand the physics, and the relevance of K-theory, let us
consider a system of $m$ D$p$-branes and $n$ anti-D$p$-branes all
of which are coincident. The gauge field configuration of the
D-branes is described by a rank $m$ vector bundle $E$ and that of
the anti-D-branes by a rank $n$ vector bundle $F$. It is natural
to postulate that complete annihilation of the branes and the
anti-branes is possible if and only if $E$ is topologically
equivalent to $F$. This requires that $m = n$, but that is not
sufficient by itself.

To be specific, let us consider equal numbers of coincident
D9-branes  and anti-D9-branes in the IIB theory for a
10-dimensional spacetime $X$.  Note that these are
spacetime-filling branes so $Y = X$.  We can represent the system
by a pair $(E, F)$, where $E$ is the vector bundle of the D-branes
and $F$ is the vector bundle of the anti-D-branes.  Next, we
define equivalence of pairs of bundles by
\begin{equation}
(E,F) \sim (E \oplus H, F \oplus H).
\end{equation}
Physically, this corresponds to adding brane-antibrane pairs with
vector bundles $(H,H)$. These branes can completely annihilate,
and therefore adding them does not change any of the charges
carried by the D-brane system. The equivalence relation described
above defines equivalence classes $[(E,F)]$. Such an equivalence
class is a K-theory class, and they form an additive group, called
the K-theory group $K(X)$. Thus we learn that the conserved
charges of type IIB D-branes are classified by $K(X)$. The
possible values of the conserved charges are in one-to-one
correspondence with the K-theory classes.

There is a similar, but somewhat different, construction for the
type IIA theory \cite{PH}, which I will not review here.  However, I would
like to remind you that in the strong coupling limit the IIA
theory becomes M theory, so all these objects should have suitable
11-dimensional counterparts.  The precise way in which this works
is a quite technical business, which was recently analyzed by
Diaconescu, Moore, and Witten \cite{DMW}.  The fact that everything works out
properly provides additional nontrivial evidence for the
consistency of the whole picture.  The group $E_8$ enters the
eleven-dimensional analysis in a surprising way, whose physical
significance is unclear.

\section{Tachyon Condensation}

Consider a spacetime-filling D25-brane in the $d=26$ bosonic
string theory.  It has a tension $T_{25} = C/g_s$, where $C$ is a
known constant.  Its dynamics is described by the bosonic open
string field theory that Witten constructed long ago \cite{W3}.  I won't
attempt to describe that theory here, but simply point out that it
is cubic in the string field $\Phi$, which is a functional of the
string embedding functions $x^\mu (\sigma)$ and world-sheet ghost
fields.  The lowest mode of $\Phi$ is a tachyon $t(x)$, so the
purely $t$-dependent part of the potential has the form $V(t) = -
\alpha t^2 + \beta t^3$ with a local maximum at $t = 0$ and a
local minimum at $t = t_0 > 0$.  Note that $V(0) = 0$ and $V(t_0)
= V_{min} < 0$.  The string field theory contains infinitely many
other scalar fields besides $t$, and it is really the complete
potential whose extremum $V_{min}$ is relevant.  The calculation
based on the cubic expression $V(t)$ only gives a first
approximation to the exact result.

We can now state the basic conjecture, due to Sen \cite{AS2}:  {\it The
minimum of the potential, where the tachyon has condensed and
other scalar fields have also adjusted appropriately, corresponds
to pure vacuum.} Therefore
\begin{equation}
T_{25} + V_{min} = 0.
\end{equation}
This means that the D-brane has completely disappeared.  One has a
configuration identical to what one would have if the D-brane
hadn't been introduced in the first place!

Open string field theory is much too complicated to carry out
analytic calculations.  So, Sen and Zwiebach \cite{SZ} tested Sen's
conjecture numerically using a level truncation approximation
scheme introduced by Kostelecky and Samuel \cite{KS}.  At leading order (the
approximation based on the cubic expression in $t$ described
above) they found a value of $V_{min}$ that is roughly 70\% of the
predicted answer.  At the level 2, where some additional
scalars and interactions are taken into account, they obtained
about 95\% of the predicted answer, and at level 4 they found 98.6\%.  
Following that, Moeller and
Taylor \cite{MT} included fields up to level 10, and all the associated
interaction terms that the approximation scheme calls for at this
order, and found 99.91\% of the expected answer.  The results are
convincing evidence for the correctness of the conjecture, even
though it is not completely clear why this approximation scheme
converges so well.

Berkovits, Sen, and Zwiebach have done the analogous calculation
for the unstable D9-brane in the type IIA theory \cite{BSZ}, which
is predicted to disappear when the tachyon condenses in exactly
the same way.  They used a version of open NS-sector superstring
field theory developed by Berkovits \cite{NB}.  (It avoids certain
technical problems associated with picture-changing operators.)
They obtained about 60\% of the expected answer at leading order
and 85\% at the next order. So this seems to be working well, too.
This calculation also applies to the annihilation of a D9 and an
anti-D9-brane in type IIB \cite{AS3}.

This D-brane disappearance story raises some interesting
questions.  The basic ones are: what has happened to the original
$U(N)$ gauge symmetry, and what has happened to the entire open
string spectrum?  The answer to both of these is that they are
confined.  What this means in terms of the string field theory
calculation is that when all the scalar fields take the values
corresponding to the minimum of the potential, the coefficients of
all kinetic terms --- both of the gauge fields and of all the
massive string modes --- become zero.  This is how confinement is
realized mathematically.  There is some numerical evidence in
support of this, but further studies would be desirable.

As an aside, let me remark that the $d=26$ bosonic string theory
also has a closed string tachyon, so it is natural to speculate
about its fate as well. However, it is not clear what is the right
conjecture to test.  Zwiebach has constructed a closed
string field theory \cite{BZ}, which is much more complicated
than open string field theory.  (Zwiebach informs me that if there were
a clear conjecture to test, he and his collaborators would find the energy to
carry out the closed string field theory computations.)
Moreover, it is not clear
to me
whether a classical computation should capture the physics
correctly or if quantum effects would play an important role. 
The point is that extremizing the potential might also force the
dilaton to a particular value. What
are some possibilities?  One possibility is that the effective
potential has no minimum and there is no stable vacuum.
Alternatively, if the effective potential does have a stable
minimum, this would seem to imply that there is a negative
cosmological constant of order string scale. One could also
imagine more exotic possibilities. Even though the bosonic string
theory is surely not relevant to the real world, it would be
satisfying to understand the answer to this question.

\section{Noncommutative Field Theory}

A field theory with noncommutativity parameters $\theta^{\mu\nu} =
- \theta^{\nu\mu}$ can be defined as a deformation of an ordinary
field theory in which field multiplication is replaced by a
nonlocal star product:
\begin{equation}
A(x) \cdot B(x) \rightarrow A * B (x),
\end{equation}
where the definition of the star product is
\begin{equation}
A * B (x) = e^{\frac{i}{2} \theta^{\mu\nu} \partial_{\mu}
\partial'_{\nu}} A (x) B (x') |_{x' = x}.
\end{equation}
The star product (also called a Moyal product) is associative but
noncommutative. In particular,
\begin{equation}
x^\mu * x^\nu - x^\nu * x^\mu = i \theta^{\mu\nu}.
\end{equation}
Thus one has a sort of Heisenberg uncertainty principle for
spacetime coordinates, and one sometimes speaks of noncommutative
geometry. Combining the two uncertainty principles gives a UV/IR
connection, which is a frequent theme in modern string theory.

Following important earlier work by Connes, Douglas, and A.
Schwarz \cite{CDS}, as well as others, Seiberg and Witten \cite{SW} gave a very clear
description of how noncommutative gauge theory arises on D-branes
when there is a nonvanishing constant $B^{(2)}$ field in the bulk.
I will not present the formulas, but simply assert that there is a
rule for getting from $B_{\mu\nu}$ in the bulk to
$\theta^{\mu\nu}$ on the brane.  (In the limit that $B$ is large
it is just the reciprocal matrix.)

On the brane, $\theta^{\mu\nu}$ corresponds to a background
Maxwell field strength.  Thus, if it has only nonzero space-space
components, this corresponds to a magnetic field on the D-brane
and describes nonlocal behavior.  The Hamiltonian still generates
unitary, causal time evolution in the usual way. Such a theory may
be useful for analyzing the quantum Hall effect.

If $\theta^{\mu\nu}$ has nonzero space-time components this
corresponds to an electric field on the brane.  This results in a
non-unitary (and hence inconsistent) field theory \cite{GM}.  In the string
theory context various authors have argued that electric fields
are okay provided they do not exceed a critical value $E_{crit}$.
This has a simple intuitive explanation: The open string has
opposite electric charges on its two ends, so when it is placed in
an electric field it gets stretched. The critical field is reached
when the stretching force balances the restoring force due to the
string tension.  Clearly, larger fields would lead to instability.

Let us now focus on the magnetic case, assuming that only
$\theta^{12} = \theta$ is nonzero.  Then, following Gopakumar,
Minwalla, and Strominger \cite{GMS}, let us consider a scalar field theory
with this noncommutativity parameter. Letting $z = x_1 + i x_2$,
and assuming no dependence on any other coordinates, the energy of
a configuration is proportional to
\begin{equation}
\int d^2 z (\partial_z \phi \partial_{\bar z} \phi + V(\phi)),
\end{equation}
where
\begin{equation}
V (\phi) = \frac{1}{2} m^2 \phi * \phi + \frac{1}{3} \lambda \phi * \phi * \phi
+ \ldots.
\end{equation}
Rescaling $z \rightarrow z \sqrt{\theta}$ gives
\begin{equation}
 \int d^2 z (|\partial \phi|^2 + \theta V (\phi)),
\end{equation}
where $*$ is now defined with $\theta = 1$.  Thus, for large
$\theta$, the extrema are given simply by
\begin{equation}
\frac{dV}{d\phi} = 0.
\end{equation}

Ordinarily, the solutions of $\frac{dV}{d\phi} = 0$ would just
give the extrema of $V$.  However, when $V$ is defined using star
products, there can be nontrivial solitonic solutions. The reason
for this is that the equation
\begin{equation}
\phi_0 * \phi_0 (x) = \phi_0 (x),
\end{equation}
has nontrivial solutions.  For example, as one can readily verify, one solution
is
\begin{equation}
\phi_0 (x) = 2e^{-(x_{1}^{2} + x_{2}^{2})}.
\end{equation}
This ``wave function'' corresponds to the ground-state projection operator
$|0\rangle\langle 0|$ of the harmonic oscillator algebra $[z, \bar z] = 1$.
Clearly, any projection operator would solve the same equation.

Using such a projection operator wave function $\phi_0$ one can
construct soliton solutions $t = t_0 \phi_0 (x)$ to $\frac{dV}{dt}
= 0$, in the notation of the previous section.  The intuitive idea
is that the core of the soliton near $x_1 = x_2 = 0$ is in the
false vacuum $t = 0$, whereas for large $x_1^2 + x_2^2$ it
approaches the true vacuum $t = t_0$.  The amazing fact about this
construction is that it only requires knowing the extrema of
$V(t)$ and not the precise form of the function that interpolates
between them.  If one accepts Sen's conjecture, this is precisely
what we do know.

Harvey et al. \cite {HKLM} have
applied these methods to bosonic open string field theory in the
manner indicated above. (Related ideas for superstrings were discussed by
Dasgupta et al. \cite{DMR}.)
They demonstrated that the ground state
$\phi_0$ soliton on a D25-brane describes a D23-brane
(concentrated near $x_1 = x_2 = 0$) with exactly the correct
tension.  This construction can be iterated to give the
D$(25-2n)$-brane. Moreover, these branes were shown to contain
tachyons with exactly the correct (imaginary) mass. The constant
$B$ field responsible for $\theta$ is pure gauge in the bulk,
where the D25-brane has disappeared as a result of tachyon
condensation.  This suggests that even though the analysis was
carried out in the limit of large $\theta$, the result should be
valid for any value of $\theta$ including $\theta = 0$.  
Harvey, et al. have found many other
interesting results, and also Witten has made some clarifying
remarks \cite{W4}.

\section{Concluding Remarks}

We have shown that studies of D-branes have led to a deeper
understanding of many issues in string theory.  As you are
probably aware, they have also had a big impact on subjects that I
have not mentioned ranging from black hole entropy to
nonperturbative properties of gauge theories.

One lesson we can learn from these developments is that open
string field theory is important because it describes D-branes.
When it was developed, this connection was not understood, and it
was not clear what it is good for. There are some indications that
to achieve background independence it will be necessary to
consider it in the $N\to \infty$ limit \cite{W5}. If a nice way to define
such a limit is found, it might prove to be very important.

Noncommutative field theory is a deformation of ordinary field
theory, which apparently does not destroy quantum consistency (in
the magnetic case). This is a remarkable fact, which has spurred a
lot of activity. Here I have focused on its use as a technical
trick for analyzing certain issues associated to tachyon
condensation on unstable D-brane systems.

As I hope to have convinced you, there have been many interesting
and fruitful developments in recent years. I am confident that
there are many more surprises to be uncovered in the coming years.
We are still quite far from a complete understanding of this
marvelous mathematical edifice.


\begin{thebibliography}{99}

\bibitem{GSW} M.B. Green, J.H. Schwarz, and E. Witten, {\it Superstring Theory},
in 2 vols., Cambridge Univ. Press, 1987.


\bibitem{JP} J. Polchinski, {\it String Theory}, in 2 vols., Cambridge Univ. Press, 1998.

\bibitem{JPD} J. Polchinski, {\it Phys. Rev. Lett.} {\bf 75} (1995) 4724, hep-th/9510017.

\bibitem{JM} J. Maldacena, {\it Adv. Theor. Phys.} {\bf 2} (1998) 231, hep-th/9711200.

\bibitem{GKP} S.S. Gubser, I.R. Klebanov, and A.M. Polyakov, {\it Phys. Lett.}
{\bf B428} (1998) 105, hep-th/9802109.

\bibitem{W1} E. Witten, {\it Adv. Theor. Math. Phys.} {\bf 2} (1998) 253, hep-th/9802150.

\bibitem{MAGOO} O. Aharony, S.S. Gubser, J. Maldacena, H. Ooguri, and Y. Oz,
{\it Phys. Rept.} {\bf 323} (2000) 183.

\bibitem{MM} R. Minasian and G. Moore, {\it JHEP} {\bf 9711} (1997) 002, hep-th/9710230.

\bibitem{W2} E. Witten, {\it JHEP} {\bf 9812} (1998) 019, hep-th/9810188.

\bibitem{PH} P. Horava, {\it Adv. Theor. Math. Phys.} {\bf 2} (1999) 373, hep-th/9812135.

\bibitem{DMW} E. Diaconescu, G. Moore, and E. Witten, hep-th/0005090 and hep-th/0005091.

\bibitem{W3} E. Witten {\it Nucl. Phys.} {\bf B276} (1986) 291.

\bibitem{AS2} A. Sen, {\it JHEP} {\bf 9806} (1998) 007, hep-th/9803194;
{\it JHEP} {\bf 9808} (1998) 010, hep-th/9805019; {\it Int. J.
Mod. Phys.} {\bf A14} (1999) 4061, hep-th/9902105.

\bibitem{SZ} A. Sen and B. Zwiebach, {\it JHEP} {\bf 0003} (2000) 002, hep-th/9912249.

\bibitem{KS} V.A. Kostelecky and S. Samuel, {\it Phys. Lett.} {\bf B207} (1998)
169; {\it Nucl. Phys.} {\bf B336} (1990) 1289.

\bibitem{MT} N. Moeller and W. Taylor, hep-th/0002237.

\bibitem{BSZ} N. Berkovits, A. Sen and B. Zwiebach, hep-th/0002211.

\bibitem{NB} N. Berkovits, {\it Nucl. Phys. } {\bf B450} (1995) 90, hep-th/9503099.

\bibitem{AS3} A. Sen, {\it JHEP} {\bf 9808} (1998) 012,
hep-th/9805170.

\bibitem{BZ} B. Zwiebach, {\it Nucl. Phys.} {\bf B390} (1993) 33, hep-th/9206084.

\bibitem{CDS} A. Connes, M.R. Douglas, and A. Schwarz, {\it JHEP} {\bf 9802} (1998) 003,
hep-th/9711162.

\bibitem{SW} N. Seiberg and E. Witten, {\it JHEP} {\bf 9909} (1999) 032,
hep-th/9908142.

\bibitem{GM} J. Gomis and T. Mehen, hep-th/0005129.

\bibitem{GMS} R. Gopakumar, S. Minwalla, and A. Strominger,
{\it JHEP} {\bf 0005} (2000) 020,
hep-th/0003160.

\bibitem{HKLM} J. Harvey, P. Kraus, F. Larsen, and E. Martinec, hep-th/0005031.

\bibitem{DMR} K. Dasgupta, S. Mukhi, and G. Rajesh,
hep-th/0005006.

\bibitem{W4} E. Witten, hep-th/0006332.

\bibitem{W5} E. Witten, talk at Strings 2000, http://feynman.physics.lsa.umich.edu/cgi-bin/s2ktalk.cgi?witten.

\end{thebibliography}
\end{document}